\documentclass{aa}
\usepackage{graphicx}
\usepackage{txfonts}
\usepackage{natbib}
\usepackage{color}
\usepackage[colorlinks, allcolors=blue]{hyperref}
\usepackage{soul}

\begin{document}

\title{Investigating datasets with high IRIS burst prevalence}

\author{C. J. Nelson$^{1}$\thanks{ESA Research Fellow}, L. Kleint$^{2,3}$}

\offprints{chris.nelson@esa.int}
\institute{$^1$European Space Agency (ESA), European Space Research and Technology Centre (ESTEC), Keplerlaan 1, 2201 AZ, Noordwijk, The Netherlands.\\
$^2$University of Geneva, 7, route de Drize, 1227 Carouge, Switzerland.\\
$^3$Astronomical Institute of the University of Bern, Sidlerstrasse 5, 3012 Bern, Switzerland.}

\date{}

\abstract
{Approximately $0.01$ \%\ of all \ion{Si}{IV} $1394$ \AA\ spectra sampled in 2013 and 2014 by the Interface Region Imaging Spectrograph (IRIS) have IRIS burst profiles. However, these events are not evenly distributed across datasets with $19.31$ \%\ of these spectra being identified in only six (of more than $3500$) rasters.}
{Here, we investigate five of these six datasets, with the aim of understanding why they contain so many IRIS burst profiles. This research will help guide future targeted analyses of IRIS bursts.}
{We analyse five datasets sampled by the IRIS satellite, studying both \ion{Si}{IV} $1394$ \AA\ spectra and $1400$ \AA\ filter slit-jaw imager (SJI) data. IRIS burst profiles are identified through the use of an automated algorithm. Additionally, we study co-spatial line-of-sight photospheric magnetic field maps sampled by the Solar Dynamics Observatory's Helioseismic and Magnetic Imager (SDO/HMI) instrument.}
{The majority of identified IRIS burst profiles ($12401$ out of $13904$) found in the five datasets analysed here were localised to seven small regions in the time-distance domain (temporal durations of $<4$ hours and spatial lengths of $<12$\arcsec\ along the slit). The SJI data co-spatial to these regions contained long-lived or repetitive compact brightenings, matching the defined properties of UV bursts, which remained close to the IRIS slit throughout their evolutions. The IRIS burst profiles were not limited to the brightest pixels in the fields of view (FOVs) nor did they comprise the majority of bright ($>500$ DN/s) pixels. These IRIS burst profiles occurred co-spatial to evolving (e.g. cancelling) opposite polarity magnetic fields where magnetic reconnection is thought to be possible.}
{More than $10$ \%\ of the IRIS burst profiles identified during the entirety of $2013$ and $2014$ are contained in just seven small regions in the time-distance domain where long-lived (lifetimes $>10$ minutes) or repetitive UV bursts occurred along the axis of the IRIS slit. }

\authorrunning{Nelson \& Kleint}
\titlerunning{Datasets with high IRIS burst prevalence}

\maketitle

\section{Introduction}
\label{Introduction}

Localised transient brightening events have been detected throughout the solar atmosphere, from the photosphere to the corona. These events are known by a variety of names, often being split into different categories based on their observational characteristics (i.e. spectral visibility and location on the Sun). In the transition region, such localised transient brightenings come in a number of forms, from explosive events (e.g. \citealt{Dere89}) to blinkers (for example, \citealt{Harrison97}), which all have unique defining spectral properties (see, for example, \citealt{Huang19}). Despite this, many of these events share several fundamental observational properties, meaning we are often able to group them together into one overarching family of UV bursts, as discussed by \citealt{Young18}. The review of these events by those authors defined five key observational properties that make a localised transient brightening in imaging data a UV burst, namely that the event is: i) of small-scale (with diameters typically below $2$\arcsec); ii) short-lived (lifetimes typically less than 10 minutes, but sometimes longer than one hour); iii) bright compared to the background atmosphere; iv) dynamic in nature, but not moving at large (typically $<10$ km s$^{-1}$) velocities across the solar disk; and v) not linked to larger-scale flares. 

The detection of localised brightenings in \ion{Si}{IV} maps, typically sensitive to plasma with temperatures around $80000$ K, by \citet{Peter14} was one of the key initial results of the Interface Region Imaging Spectrograph (IRIS; \citealt{dePontieu14}). One of the most interesting aspects of these spectra, now commonly referred to as IRIS bursts, was that the \ion{Si}{IV} $1394$ \AA\ profiles contained strong absorption lines from cooler transitions, indicating that they potentially formed deep in the solar atmosphere. This interpretation has, however, been questioned in the literature by \citet{Judge15}, meaning that future work is still required. As IRIS bursts were found close to inversion lines between opposite polarity magnetic fields (see, for example, \citealt{Peter14, Nelson16}), it was hypothesised that these events could be driven by magnetic reconnection, which was heating the local plasma to temperatures an order of magnitude higher than typically found in the photosphere. The co-spatial and co-temporal occurrence of IRIS bursts with photospheric bursts, such as Ellerman bombs (EBs; \citealt{Ellerman17}) which have typical temperatures below $10^4$ K, as reported by \citet{Vissers15} and \citet{Tian16}, further supported the theory that these events are driven by magnetic reconnection. How such magnetic reconnection would cause responses across this large range of temperatures is still not understood though (see, for example, \citealt{Reid17}). The fact that IRIS burst profiles could form at temperatures as low as $15000$ K could go some way to explain this issue (\citealt{Rutten16}), with it also being possible that magnetic reconnection is occurring co-temporally at different heights (offset by several hundreds of km) in the solar atmosphere causing these varied signatures (see e.g. \citealt{Hansteen17, Chen19}). Crucially, not all IRIS burst profiles are found co-spatial to features that match the definition of UV bursts in imaging data.

In order to investigate the importance of IRIS bursts towards answering global questions in solar physics (e.g. coronal heating), \citet{Kleint22} developed an algorithm (using techniques defined in \citealt{Panos21a, Panos21b}) that could identify whether any given \ion{Si}{IV} $1394$ \AA\ spectra matched the observed properties of such events. In particular, the algorithm searched for pixels whose \ion{Si}{IV} $1394$ \AA\ spectra also contained super-imposed absorption lines, as in the original examples presented by \citet{Peter14}. No specific intensity threshold was applied to the spectra, in order to identify as wide a sample of events as possible. Those authors studied all IRIS \ion{Si}{IV} $1394$ \AA\ spectra sampled throughout 2013 and 2014, finding that approximately $0.01$ \%\ could be classed as IRIS burst profiles. This work was followed up by \citet{Nelson22} who made several minor modifications to the algorithm (such as removing the criterion that $30$ spectra of interest must be present in a dataset for it to be studied further) in order to study whether the properties (e.g. frequencies, areas, locations, spectral shapes) of IRIS bursts varied as active regions (ARs) evolved.  It was found that the observable characteristics of these spectra were relatively stable as their host ARs aged, changing only slightly between datasets without any predictable or consistent pattern. This research indicated that the number of IRIS bursts in an AR may not be useful as a predictor of larger-scale features (e.g. flares) at a later time. The interested reader should consult \citet{Kleint22} and \citet{Nelson22} for more information about the algorithm itself.

Another key result which was obtained from the large-scale statistical study of \citet{Kleint22} was that IRIS burst profiles are not evenly distributed throughout the IRIS data catalogue. Specifically, those authors studied $3537$ IRIS datasets with more than $30$ candidate IRIS burst profiles being detected in fewer than $10$ \%\ of these. Indeed, one dataset within the sample of \citet{Kleint22} contained $4.77$ \%\ of all IRIS burst profiles recorded in $2013$ and $2014$, whilst no IRIS burst profiles were detected in datasets sampling the quiet-Sun. An important implication of this is that if magnetic reconnection is responsible for heating the quiet corona then it is occurring with properties which would cause transient brightenings with different spectral signatures to IRIS bursts, such as campfires (\citealt{Berghmans21}) observed by the Solar Orbiter satellite (\citealt{Muller20}). In addition to this, \citet{Nelson22} found that not all IRIS datasets sampling ARs contained IRIS burst profiles either. It is, therefore, of interest to examine those datasets where a large number of IRIS burst profiles were identified in order to investigate whether the physical conditions at these locations are in any way unique.

\begin{figure}
\includegraphics[width=0.49\textwidth]{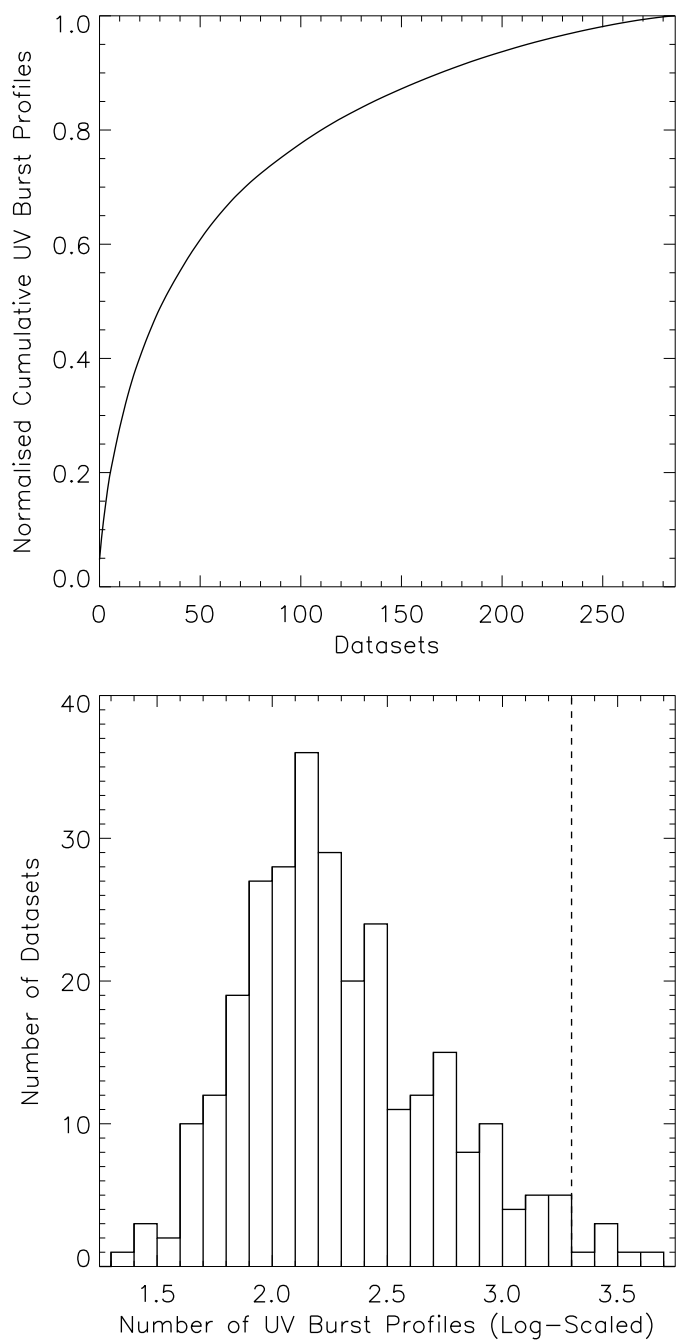}
\caption{General statistics of IRIS burst profiles from the sample of \citet{Kleint22}. (Top panel) CDF for the total number of IRIS burst profiles across all $287$ datasets found to contain such spectra, clearly demonstrating that these events are not evenly distributed. (Bottom panel) Histogram plotting the number of datasets containing specific numbers of IRIS burst profiles. The vertical dashed line indicates the cut-off used here to define high IRIS burst prevalence, with five of the six datasets to the right being studied in this article.}
\label{UVMax_fig1}
\end{figure}

In this article, we use the algorithm developed by \citet{Kleint22} and \citet{Nelson22} to investigate the properties of IRIS burst profiles, as well as the background atmosphere, in datasets found to contain an extremely large numbers of such spectra. This research will help to guide future work into IRIS burst profiles using the IRIS data catalogue after $2014$. Our work is set out as follows: In Sect.~\ref{Methods} we detail our data selection criteria and outline the basic properties of the datasets analysed here; In Sect.~\ref{Results} we present the specific datasets studied and the results of our analysis; before in Sect.~\ref{Conclusions} we include a brief summary and draw our conclusions.

\section{Methods}
\label{Methods}

\begin{table*}
\caption{General information about the five datasets studied in this article.}
\begin{center}
\begin{tabular}{| c | c | c | c | c | c | c | c | c | c | c | c | c | c |}
\hline
\bf{Date} & \bf{Start} & \bf{End} & \bf{n$_\mathrm{step}$} & \bf{n$_\mathrm{rast}$} & \bf{t$_\mathrm{exp}$} & \bf{t$_\mathrm{cad}$} & \bf{p$_\mathrm{slit}$} & \bf{FOV$_\mathrm{x}$} & \bf{FOV$_\mathrm{y}$} & \bf{$\Delta\lambda$} & \bf{B$_\mathrm{KP}$} & \bf{B$_\mathrm{NK}$} & \bf{OBSID} \\ 
 & \bf{(UT)} & \bf{(UT)} & & & \bf{(s)} & \bf{(s)} & & \bf{(\arcsec)} & \bf{(\arcsec)} & \AA\ & & & \\ \hline
2013-08-31 & 07:49:35 & 11:39:57 & 2 & 2280 & 2 & 6 & 362 & 1 & 51 & 0.025 & 3134 & 2099 & 4183254165 \\ \hline
2013-12-06 & 19:10:41 & 19:50:14 & 4 & 120 & 4 & 20 & 396 & 1 & 60 & 0.025 & 2705 & 2230 & 3800256165 \\ \hline
2014-03-09 & 11:29:52 & 21:39:03 & 1 & 2250 & 15 & 16 & 401 & 0 & 60 & 0.050 & 2300 & 1961 & 3820510152 \\ \hline
2014-09-07 & 11:24:59 & 17:59:20 & 1 & 2500 & 8 & 10 & 775 & 0 & 119 & 0.025 & 4836 & 4434 & 3820259253 \\ \hline
2014-09-18 & 08:02:53 & 10:09:08 & 1 & 1400 & 4 & 6 & 775 & 0 & 119 & 0.025 & 3805 & 3190 & 3820257453 \\ \hline
\end{tabular}
\end{center}

\tablefoot{Columns denote: The date of the observation; the start time of the observation; the end time of the observation; the number of raster steps; the number of raster repeats; the exposure time per raster step; the total cadence of each raster; the number of pixels along the slit; the FOV of each raster in the $x$-direction; the FOV of each raster in the $y$-direction; the spectral sampling in the \ion{Si}{IV} $1394$ \AA\ window; the number of IRIS burst profiles identified by \citet{Kleint22}; the number of IRIS burst profiles identified after the modifications to the algorithm applied by \citet{Nelson22}; and the OBSID of the experiment.}
\label{UVMax_tab1}
\end{table*}

In order to identify datasets with a high prevalence of IRIS burst profiles, we initially studied the general properties of all $287$ IRIS datasets reported to contain such spectra by \citet{Kleint22}. In the top panel of Fig.~\ref{UVMax_fig1}, we plot a cumulative distribution function (CDF) for the spread of IRIS burst profiles across these $287$ datasets. Clearly a small number of datasets contained a large proportion of the IRIS burst profiles, with the most populous $50$ datasets returning $60.03$ \%\ of all IRIS burst profiles. As a contrast, the most sparse $50$ datasets contained only $2.84$ \%\ of all identified IRIS burst profiles. In the bottom panel of Fig.~\ref{UVMax_fig1}, we plot a histogram of the number of IRIS burst profiles per dataset (log-scaled) against frequency. The mode of this distribution was between $2.1$-$2.2$ (corresponding to $126$-$158$ IRIS burst profiles), however, it was found that $20$ datasets contained more than $1000$ events. In order to reduce this larger sample to a manageable number of datasets for analysis, we limited our research here to datasets that contained more than $2000$ IRIS burst profiles (indicated by the vertical dashed line) and contained fewer than $10$ raster steps. The five datasets that satisfied these criteria contained $16.56$ \%\ ($16780$) of all IRIS burst profiles ($101337$) identified in the sample of \citet{Kleint22}. Notably, only one dataset was found to contain more than $2000$ IRIS burst profiles that did not match the raster step criterion, with this dataset having been studied in the context of IRIS bursts previously by \citet{Vissers15}. 

After reapplying the algorithm with the modifications described in \citet{Nelson22}, the total number of IRIS burst profiles available for analysis within the five datasets studied here reduced slightly to $13904$. To investigate the reason for this, we manually inspected a sample of the spectra rejected by this iteration of the algorithm that were previously classed as IRIS burst profiles, as well as a sample of IRIS burst profiles that were previously rejected by the algorithm but were found to be IRIS burst profiles here. We qualitatively found that the IRIS burst profiles found uniquely by the new version of the algorithm (as used here) had a reduced rate of false positive detections, when compared to those IRIS burst profiles identified uniquely by the previous version of the algorithm (as used by \citealt{Kleint22}). This indicates that the algorithm is performing well in general, but that it is very difficult to define unique criteria to find every single burst spectrum reliably whilst also avoiding false positives. We also studied the spatial locations of the potential IRIS burst profiles found uniquely by the old version of the algorithm, finding that they were predominantly grouped together closely with the IRIS burst profiles identified using the current version of the algorithm. Overall, the fact that $97.01$ \%\ of all of the IRIS burst spectra studied here were found by both versions of the algorithm gives us confidence that both previous results and those results presented here remain valid.

For each of these five datasets, we analysed both IRIS $1394$ \AA\ spectra (used to identify the locations of IRIS burst profiles) and slit-jaw imager (SJI) data. Specifically, we analysed the evolution of the wider transition region using images sampled using the \ion{Si}{IV} $1400$ \AA\ SJI filter. The spatial sampling along both the spectral slit and within the SJI field-of-view (FOV) was $0.167$\arcsec\ for each raster, however, other fundamental properties of these datasets differed. Further targeted information about each dataset (dates and times, number of raster steps and repeats, exposure times and cadences, FOVs of the rasters, spectral sampling in the \ion{Si}{IV} $1394$ \AA\ window, number of IRIS burst profiles returned using the algorithm, IRIS OBSID) is, therefore, summarised in Table~\ref{UVMax_tab1}. Finally, we also analysed the evolution of the line of sight magnetic field co-spatial and co-temporal to the detected IRIS burst profiles using data sampled by the Solar Dynamics Observatory's Helioseismic and Magnetic Imager (SDO/HMI; \citealt{Scherrer12}). These data have a post-reduction pixel scale of $0.6$\arcsec\ and a cadence of $45$ s.

\section{Results and discussion}
\label{Results}

\subsection{Structuring of IRIS burst profiles}

\subsubsection{31 August 2013}

\begin{figure*}
\includegraphics[width=0.99\textwidth]{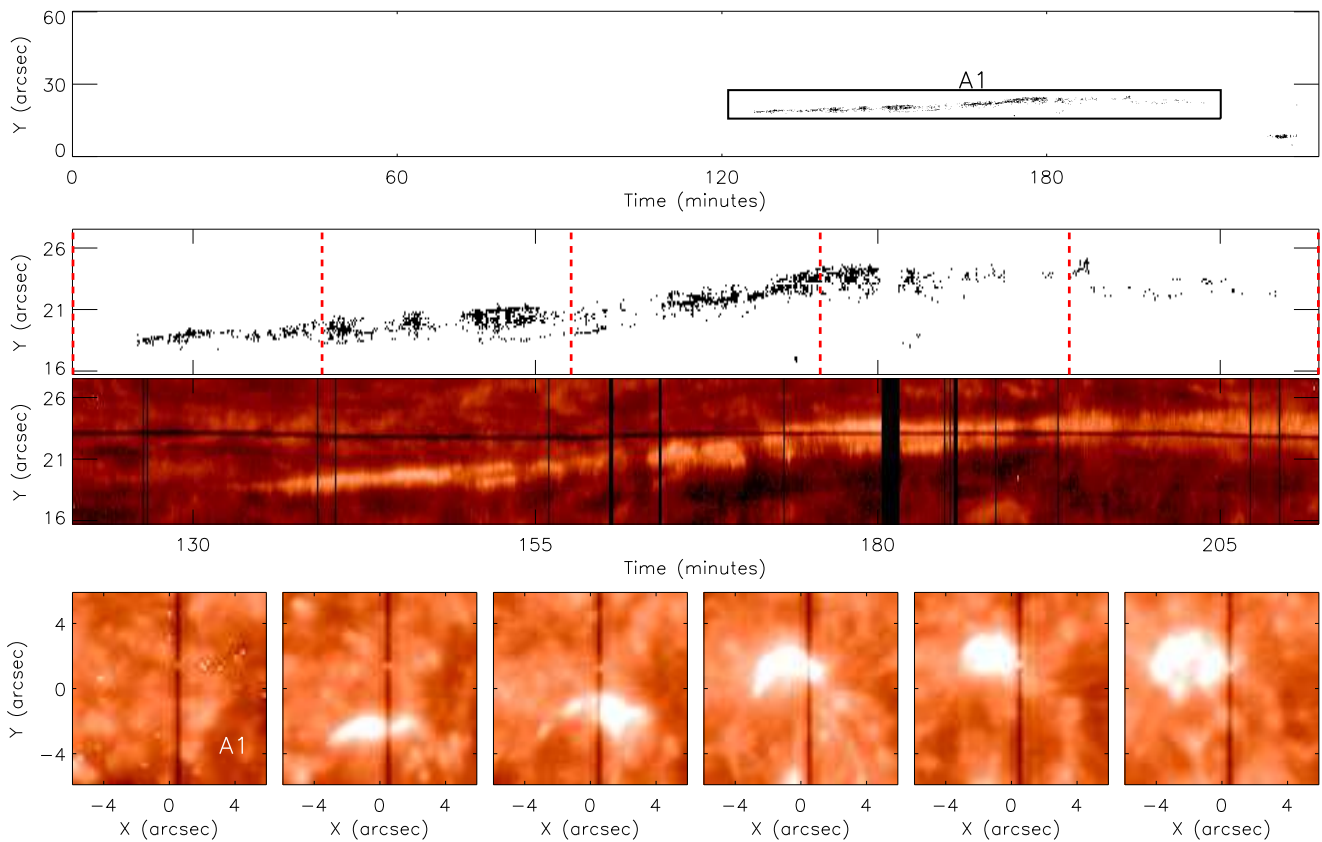}
\caption{Properties of the IRIS burst profiles within the dataset sampled on $31$ August $2013$. (Top row) Time-distance binary map for the entire dataset, with black pixels pin-pointing IRIS burst profiles identified using our algorithm. The majority ($91.57$ \%) of IRIS burst profiles are contained within a small region highlighted by the box labelled `A1'. (Second row) Time-distance binary map plotting only the region within A1. The vertical dashed red lines indicate the times of the SJI $1400$ \AA\ images plotted in the bottom row. (Third row) Time-distance intensity map (logarithmically scaled) calculated from the \ion{Si}{IV} $1394$ \AA\ line core (rest wavelength) co-spatial to the FOV plotted in the second row. The vertical black lines denote regions where no data was returned. (Bottom row) Six $1400$ \AA\ SJI frames (logarithmically scaled) sampled over the $90$ minutes covered by A1. Clearly no compact brightening is initially present (first panel), before one appears (second panel), moves upwards parallel to the IRIS slit (third to fifth panels), before moving to the left of the slit (sixth panel) such that it was no longer detectable in the spectral data.}
\label{UVMax_fig2}
\end{figure*}

\begin{figure*}
\includegraphics[width=0.99\textwidth]{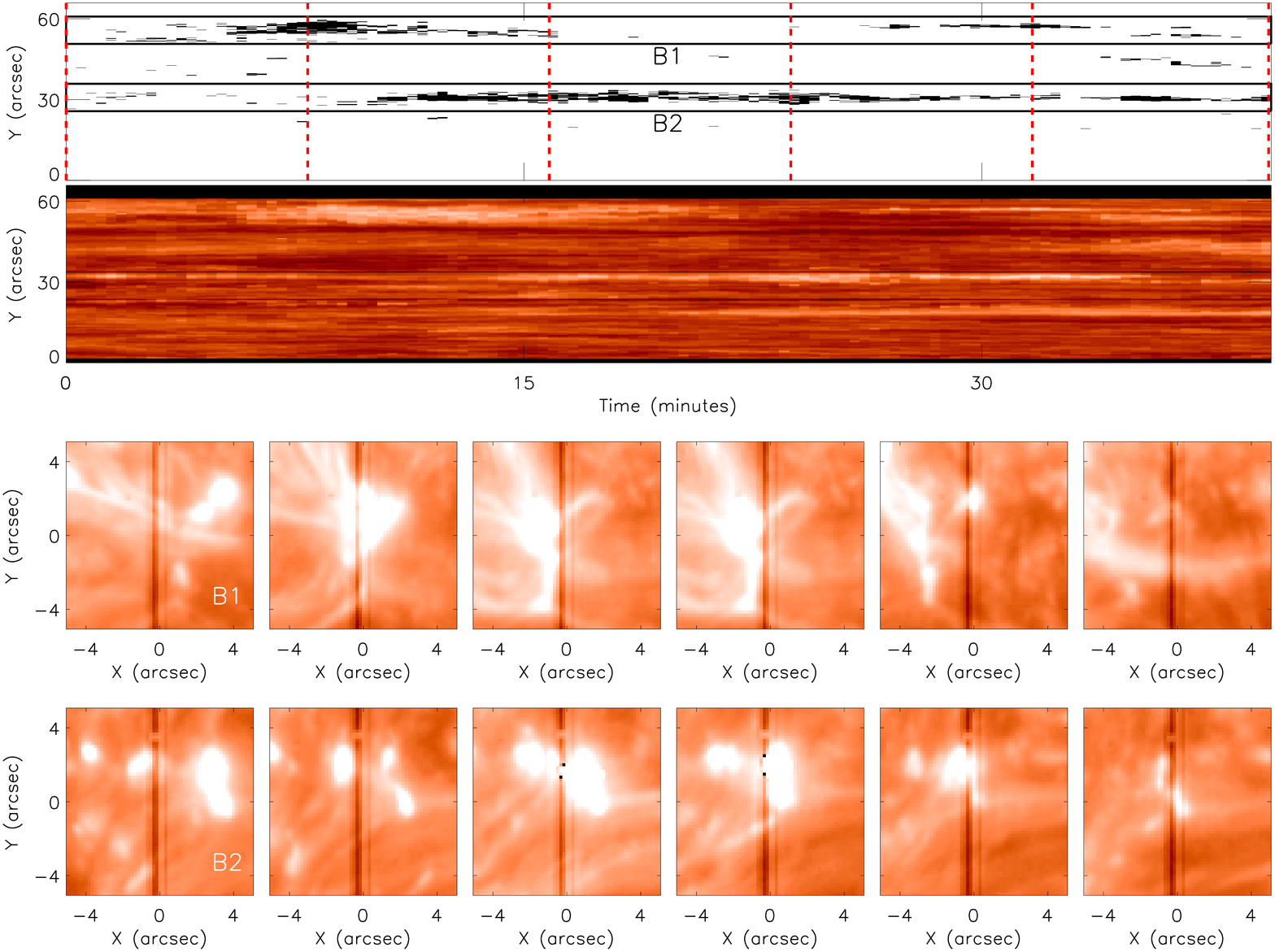}
\caption{Similar to Fig.~\ref{UVMax_fig2}, but for the dataset sampled on $6$ December $2013$. (Top row) Binary time-distance plot displaying the locations of IRIS burst profiles throughout this $40$ minute dataset. The black boxes labelled `B1' and `B2' indicate two regions of increased IRIS burst activity along the slit, whilst the vertical dashed red lines indicate the times at which the SJI data panels plotted in the bottom two panels were sampled. (Second row) \ion{Si}{IV} $1394$ \AA\ line core (rest wavelength) intensity (logarithmically scaled) co-spatial to the binary map plotted above. (Third row) Six zoomed-in FOVs from the SJI data (logarithmically scaleds) with the $y$-axis being equal to the height of B1. (Bottom row) Same as for the third row but for B2.}
\label{UVMax_fig3}
\end{figure*}

We began our analysis by studying the first dataset reported in Table~\ref{UVMax_tab1}, ordered chronologically. This dataset, sampled on $31$ August $2013$, essentially comprised of a two-step, short ($51$\arcsec\ slit length) raster that covered a plage region slightly north of a sunspot within AR $11836$ for close to four hours. During this time, $2099$ ($0.13$ \%\ of all pixels in this dataset) IRIS burst profiles were identified through our automated analysis with a temporal frequency of $0.15$ per second (number of IRIS burst profiles divided by length of observation in seconds). In the top panel of Fig.~\ref{UVMax_fig2}, we plot a binary time-distance map displaying the locations of all IRIS burst profiles within this dataset (indicated by the black pixels), identified using our automated methods. This plot is created by considering the location of the IRIS burst profile along the slit and the raster within which it was detected, with events observed at both raster steps included. Clearly, the majority of the returned events are contained within just a small region, highlighted by the over-laid black box labelled `A1'. This box covers approximately $12$\arcsec\ along the slit (indicating that the IRIS bursts are confined to a small spatial region) and a temporal duration of around 90 minutes. Specifically, $1922$ ($91.57$ \%) of the $2099$ IRIS burst profiles identified for this dataset using our algorithm are contained within A1, with the remainder of events found in a small region at the bottom right of the plot.

\begin{figure*}
\includegraphics[width=0.99\textwidth]{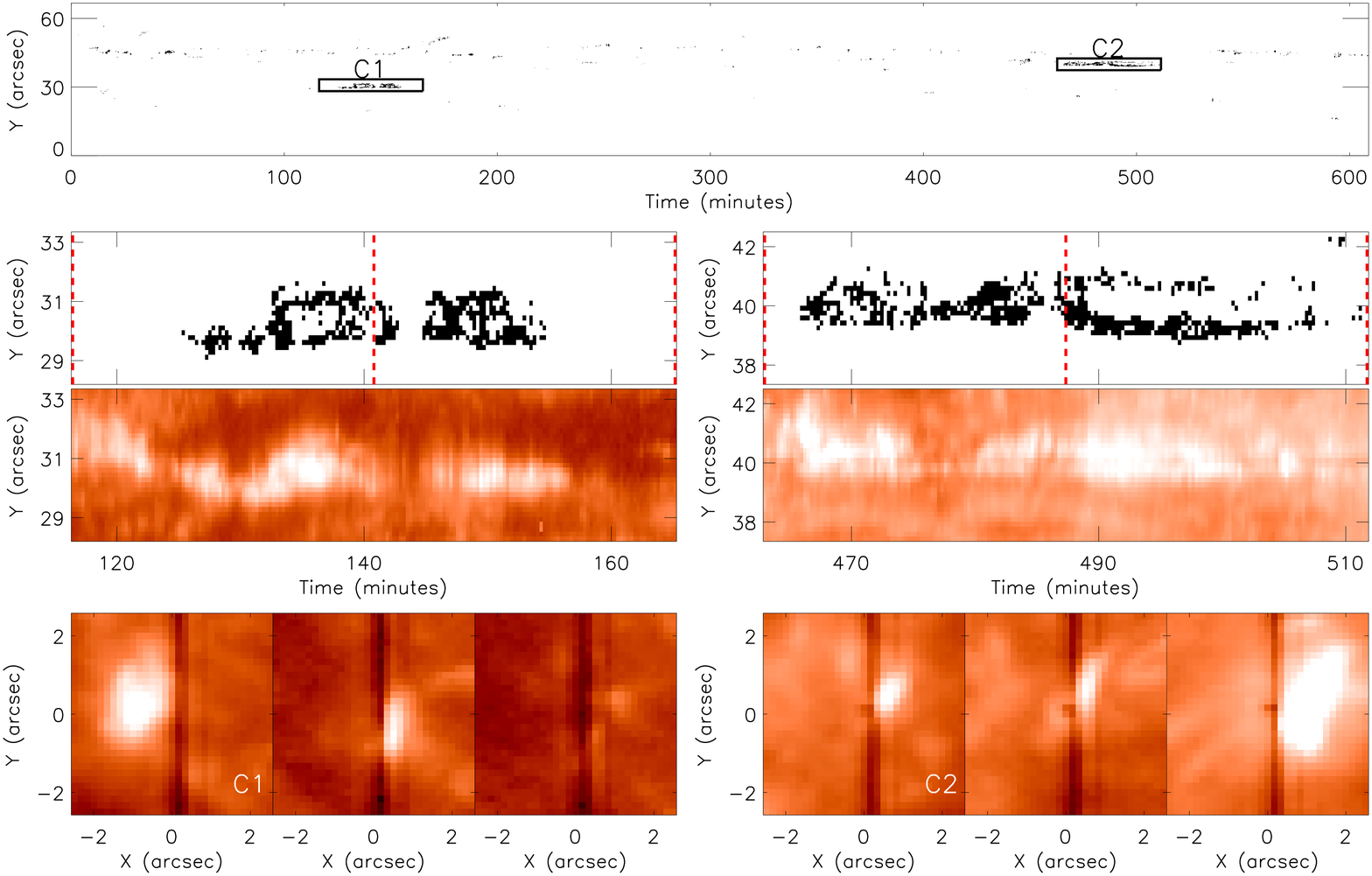}
\caption{Similar to Fig.~\ref{UVMax_fig2} but for the dataset sampled on $9$ March 2014. The left and right columns, from the second row downwards, correspond to C1 and C2, respectively.}
\label{UVMax_fig4}
\end{figure*}

In the second row of Fig.~\ref{UVMax_fig2}, we plot the same binary time-distance map as plotted in the top row but now covering only the region contained within A1. Clearly, the IRIS burst profiles appear to be well-grouped at each time-step (typically being found within a $3$\arcsec\ distance along the slit) with the centre-of-mass moving from $Y\approx18$\arcsec\ at $t\approx125$ minutes to $Y\approx24$\arcsec\ at $t\approx180$ minutes, corresponding to a plane-of-sky velocity along the slit of approximately $1.32$ km s$^{-1}$. Analysis of co-temporal SJI movies indicated that these apparent motions appeared to be of solar origin and not due to variations in IRIS' pointing through time, as occurred during some of the following data sets. The six vertical dashed red lines (including both $y$-axes) indicate the times at which the SJI data plotted in the bottom row were sampled. The third row of Fig.~\ref{UVMax_fig2} plots the \ion{Si}{IV} $1394$ \AA\ core intensity (logarithmically scaled) co-spatial to the second row, constructed using the intensity at the first raster step. The black vertical lines indicate rasters where no data was returned. Notably, the thick black line after $t$=$180$ corresponds to a gap in IRIS burst profile detection, as would be expected. We find that the IRIS burst profiles were predominantly identified close to, but not exclusively within, a small (length $<3$\arcsec) region of increased intensity compared to the background, which moved along the IRIS slit with a comparable speed to the detected IRIS burst profiles.

Finally, we analysed the evolution of plasma sampled by the SJI $1400$ \AA\ filter co-spatial to A1. In the bottom row of Fig.~\ref{UVMax_fig2} we plot a square SJI FOV at six different time-steps, each separated by approximately $18$ minutes. The $x$-axis of these panels is centred on the location of the IRIS slit and the $y$-axis is the same as that plotted in the second and third rows, shifted to the origin for ease of comparison. Initially no compact brightening was present within this FOV (first panel), however, $18$ minutes later a clear brightening had developed (second panel). This feature evolved dynamically over the next $54$ minutes (third, fourth, and fifth panels), occasionally appearing to emit short jets that propagated towards the bottom of the FOV, as it moved along the axis of the IRIS slit. At the end of this time (sixth panel), the feature moved to the left of the IRIS slit, where it continued to be present until the end of the dataset, returning an observed lifetime of close to $100$ minutes. Overall, this feature matched the list of criteria that define UV bursts (as discussed by \citealt{Young18}) well, albeit with sizes and lifetimes close to the upper limit for such events.

\begin{figure*}
\includegraphics[width=0.99\textwidth]{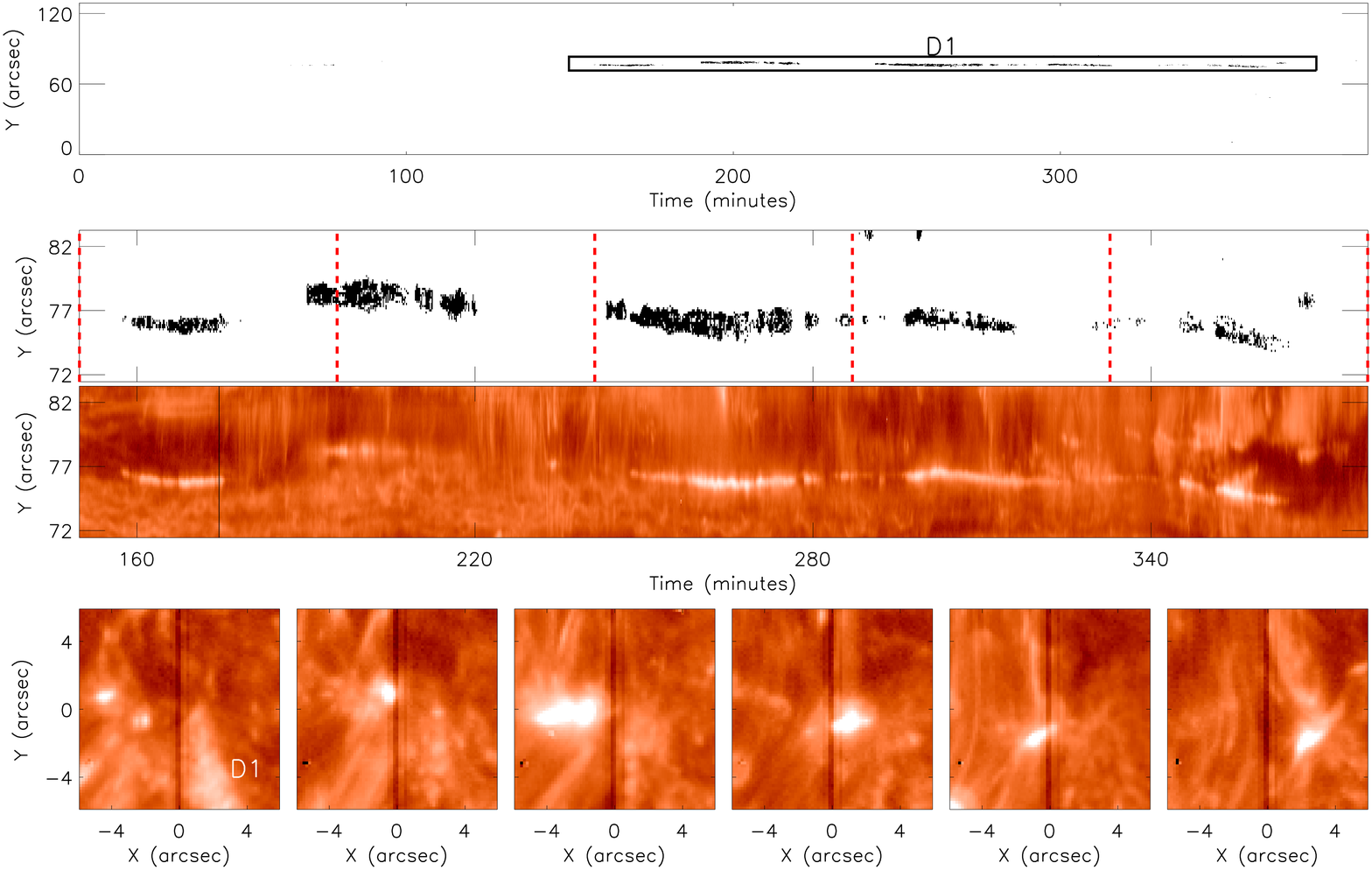}
\caption{Same as for Fig.~\ref{UVMax_fig2} but for the dataset sampled on $7$ September $2014$. The region containing the majority of IRIS burst profiles is indicated by the black box labelled `D1'.}
\label{UVMax_fig5}
\end{figure*}

\begin{figure*}
\includegraphics[width=0.99\textwidth]{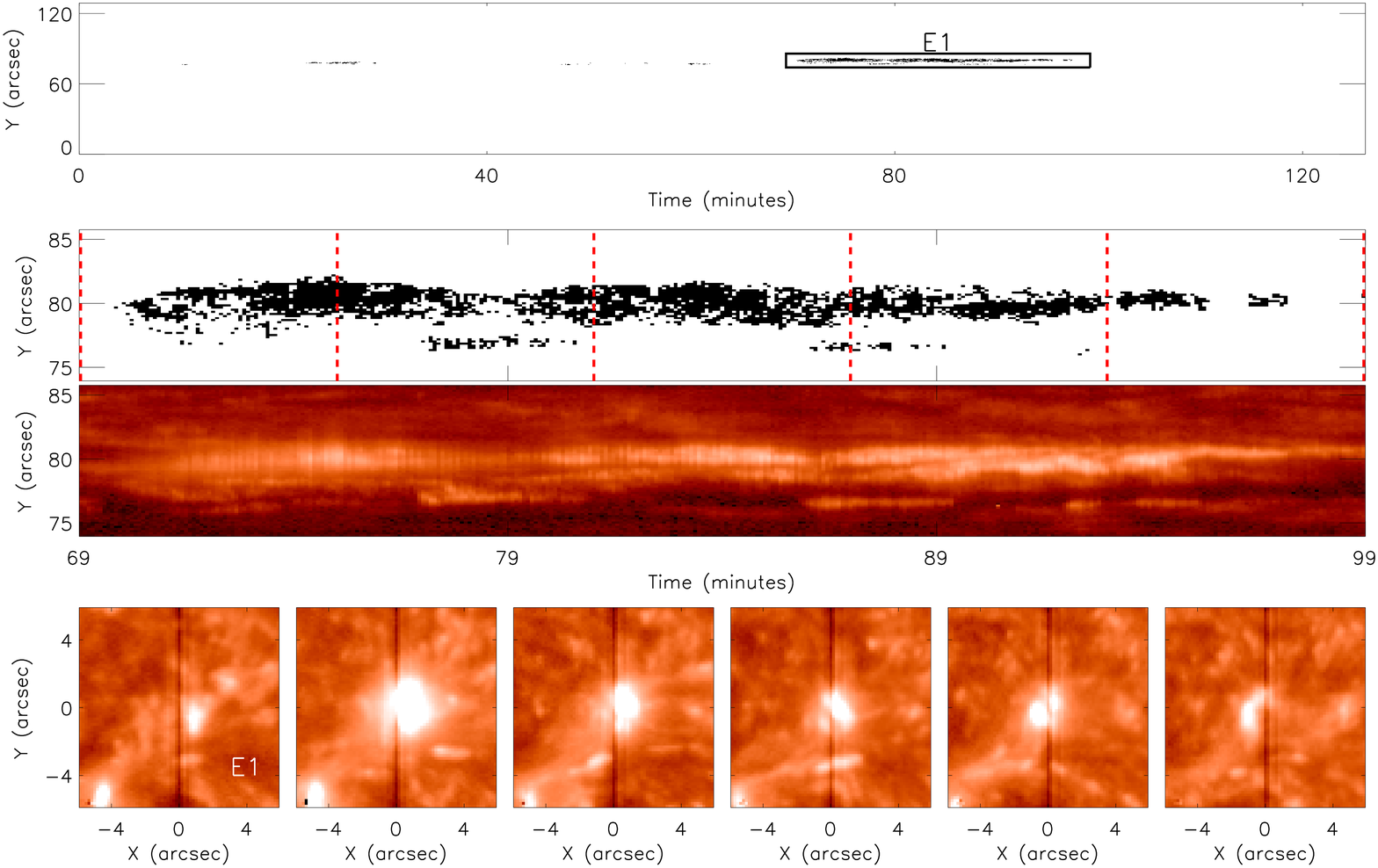}
\caption{Same as for Fig.~\ref{UVMax_fig2} but for the dataset sampled on $18$ September $2014$. The region containing the majority of IRIS burst profiles is highlighted by the black box labelled `E1'.}
\label{UVMax_fig6}
\end{figure*}

\subsubsection{6 December 2013}

The second dataset studied here consisted of a short ($60$\arcsec\ slit length), four-step dense raster sampled on $6$ December $2013$. This raster sequence observed the centre of AR $11916$, between the lead and trailing sunspots, for just under $40$ minutes, making it the shortest time series analysed in this article. Notably, AR $11916$ was found to be host to the most IRIS burst profiles of any AR studied in the sample of \citet{Nelson22}. A total of $2230$ ($1.17$ \%\ of all pixels in this dataset) IRIS burst profiles were detected during this observational experiment corresponding to a frequency of $0.93$ per second. The IRIS burst profiles within this dataset were predominantly found in two separate $10$\arcsec\ length windows, which are indicated by the over-laid black boxes labelled `B1' and `B2' on the binary distance-time map plotted in the top panel of Fig.~\ref{UVMax_fig3}. Specifically, $759$ ($34.04$ \%) of the IRIS burst profiles identified in this dataset were contained within B1, whilst $1399$ ($62.74$ \%) of the IRIS burst profiles were identified within B2. Once again, the vertical dashed red lines indicate the times at which the SJI panels (plotted in the bottom two rows) were sampled. The IRIS burst profiles identified in both of these regions displayed no apparent plane-of-sky velocities along the IRIS slit during these 40 minutes. Qualitatively comparing the locations of these IRIS bursts profiles to the \ion{Si}{IV} $1394$ \AA\ line core (rest wavelength) intensity (second row of Fig.~\ref{UVMax_fig3}; logarithmically scaled) once again indicated these spectra predominantly occurred close to, but not exclusively within regions that were bright compared to the local background.

Analysis of the SJI data revealed that dynamic compact brighenings were present co-spatial to B1 and B2 during this time. In B1 (third row of Fig.~\ref{UVMax_fig3}), a compact brightening (at co-ordinates of $x\approx3$\arcsec, $y\approx3$\arcsec) was initially evident (first panel), before this feature moved across the FOV from right to left passing over the IRIS slit (second, third, and fourth panels). This corresponds to the first grouping of IRIS burst profiles in B1 (before $t\approx15$). This feature continued to move across the FOV, until it was out of the IRIS slit, before a second compact brightening (co-ordinates of $x\approx0$\arcsec, $y\approx2$\arcsec) then occurred directly over the slit (fifth panel) contributing the second grouping of IRIS burst profiles in B1 (after $t\approx25$). This second brightening then faded away before the final frame of the dataset (sixth column). The second region, B2, also contained a wealth of activity. A compact brightening (co-ordinates of $x\approx4$\arcsec, $y\approx0$\arcsec) was initially present to the right of the FOV (first panel) before this feature moved towards the slit (second and third panels), where it remained for the next $\approx30$ minutes (fourth and fifth panels). Occasional small-scale jets were once again released towards the bottom left of the FOV. This compact brightening decreased in size but was still present at the end of the time series (sixth panel). We note that the apparent plane-of-sky motions (from right to left) evident in this FOV were predominantly caused by uncorrected wobbling from the instrument remaining in these data, rather than actual motions of these compact brightenings on the Sun. We find that these compact brightenings have observed properties consistent with larger and longer-lived UV bursts (\citealt{Young18}).

\subsubsection{9 March 2014}

The third dataset analysed in this article was sampled on $9$ March $2014$ over the course of slightly more than $10$ hours, making it the longest dataset studied here. During this time, IRIS observed a region of emerging flux within AR $11996$ in sit-and-stare mode using a short ($60$\arcsec\ in length) slit. A total of $1961$ ($0.22$ \%\ of all pixels in this dataset) IRIS burst profiles were detected in these data using our automated methods, corresponding to a temporal frequency of $0.05$ per second. In the top panel of Fig.~\ref{UVMax_fig4}, we plot a binary distance-time plot displaying the distribution of IRIS burst profiles throughout this raster. It is immediately evident that the detected IRIS burst profiles were more evenly spread than those found within the two previous datasets with a large number of one or two pixel black regions present throughout the entire time series. However, two regions (black boxes labelled `C1' and`C2') of this diagram, with lengths of $5$\arcsec\ along the slit and durations of $48$ minutes, were found to contain the majority of the total number of IRIS burst profiles. Specifically, $451$ ($23.00$ \%) of the IRIS burst profiles identified in this dataset were contained within C1, whilst $591$ ($30.14$ \%) were contained within C2.

In the second row of Fig.~\ref{UVMax_fig4}, we plot time-distance binary maps displaying the locations of IRIS burst profiles contained within C1 (left panel) and C2 (right panel). Once again, the vertical dashed red lines (of which there are three for each column) indicate the times of the SJI data plotted in the bottom row. Clearly, the IRIS burst profiles within these regions were isolated to short distances along the slit of approximately $2$\arcsec\ and were only detected for between $20$-$45$ minutes. No clear plane-of-sky motions of the IRIS burst profiles parallel to the slit were apparent. Comparing the locations of the IRIS burst profiles to maps of the \ion{Si}{IV} $1394$ \AA\ line core (rest wavelength) intensity (third row of Fig.~\ref{UVMax_fig4}; logarithmically scaled) once again revealed that these spectra were typically identified close to, but not exclusively within, regions of increased intensity.

Analysis of the SJI data revealed that the majority of IRIS burst profiles in both C1 and C2 were co-spatial to individual compact brightenings, that formed and evolved close to the IRIS slit. With respect to C1, a compact bright region was initially evident to the left of the slit (first panel), before this moved right into the axis of the slit (second panel). The compact brightening then completely faded away, to the extent that it was no longer visible around $24$ minutes later (third panel). Regarding C2, a compact brightening initially formed along the line of the slit (first panel), with this feature remaining relatively stationary for around $24$ minutes (second panel), before moving right, out of the line of the slit (third panel) where it remained until it faded away shortly after. As with the previous dataset, the apparent motions within these SJI data were predominantly caused by uncorrected wobble during this time, rather than physical motions of the IRIS bursts. We once again found that the compact brightenings in the SJI data that was co-spatial to many of these IRIS burst profiles matched well with the criteria that define UV bursts (\citealt{Young18}).

\subsubsection{7 September 2014}

The fourth dataset studied here was collected by IRIS on $7$ September $2014$. This $395$ minute observation consisted of a large ($119$\arcsec\ slit length) sit-and-stare sequence with pointing centred along an extended polarity inversion line through the centre of AR $12157$ (roll angle of around $60^\circ$). This raster was found to contain $4434$ ($0.23$ \%\ of all pixels in this dataset) IRIS burst profiles, the most of any dataset from $2013$ and $2014$, with a temporal frequency of 0.19 per second. In the top row of Fig.~\ref{UVMax_fig5}, we plot a binary time-distance map displaying the locations of the IRIS burst profiles detected during this time using our automated methods. Once again the majority of the IRIS burst profiles identified were restricted to one localised region (around $12$\arcsec\ along the slit and $228$ minutes long). This region (denoted by the black box labelled `D1') contained $4370$ ($98.56$ \%) of the IRIS burst profiles found in this dataset.

\begin{figure*}
\includegraphics[width=0.99\textwidth]{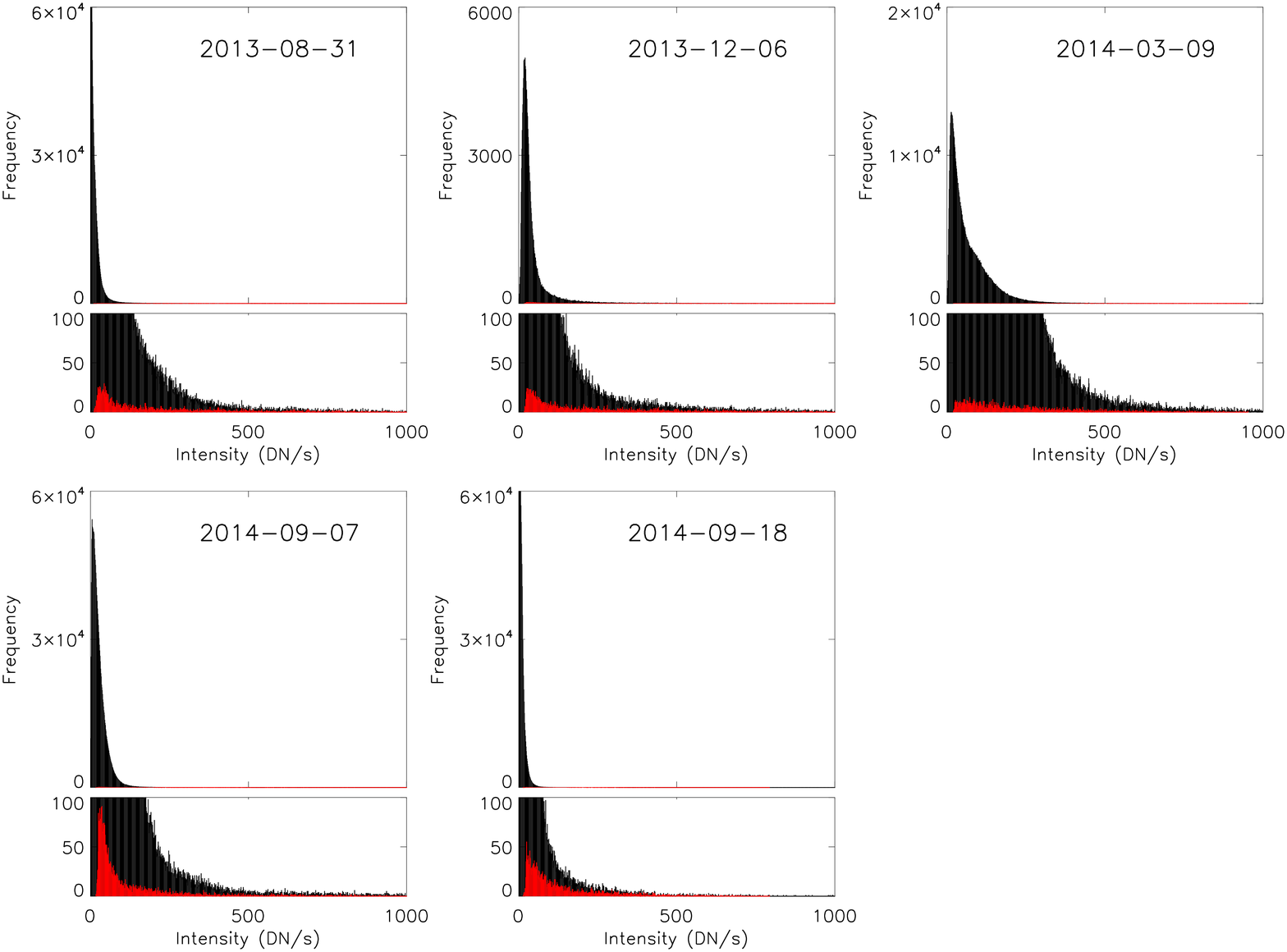}
\caption{Distributions of the \ion{Si}{IV} $1394$ \AA\ line core (rest wavelength) intensity for the five datasets studied here. (Large panels) Histograms of the intensity for each pixel within the rasters studied here (black bars), as well as the intensity only at locations where IRIS burst profiles were detected (red bars). (Small panels) Same as above but with the $y$-axis scaled to better display the red bars. Clearly, IRIS burst profiles were not confined to only the brightest pixels in the FOVs.}
\label{UVMax_fig7}
\end{figure*}

Analysing the binary time-distance map within D1 more closely (second row of Fig.~\ref{UVMax_fig5}), we found that the majority of the IRIS burst profiles were linked to several short ($20$-$40$ minutes) bouts of activity, with some of these appearing to move slowly (typically velocities of $<1.5$ km s$^{-1}$) along the axis of the slit. Given the pointing of IRIS remained relatively stable during this time (confirmed by studying the SJI movies), it is likely that these velocities are evidence of real motions (with some potential contributions from slight uncorrected wobble). Once again, the vertical dashed red lines indicate the time-steps plotted for the SJI data (bottom row). Qualitatively comparing the locations of these IRIS burst profiles to the \ion{Si}{IV} $1394$ \AA\ line core (rest wavelength) intensity (third row; logarithmically scaled), we again found that these spectra typically occurred close to regions of increased intensity, that also appeared to move along the axis of the IRIS slit at low velocities. The fact that IRIS burst profiles do not occur exclusively within bright regions was most evident at $t\approx260$, where the grouping of identified IRIS burst profiles was more than double the height of the thin bright region in the intensity map.

To further understand the dynamics within this region, we investigated the SJI data co-spatial to D1 (six time-steps are plotted in the bottom row of Fig.~\ref{UVMax_fig5}). A range of compact brightenings were present within this FOV over the course of these $228$ minutes. Initially, no compact brightenings were present (first panel), before a feature (diameter of around $1$\arcsec) appeared along the axis of the slit (corresponding to the grouping of IRIS burst profiles starting at $t\approx160$ in D1). This compact brightening then moved right, out of the axis of the slit, and faded from view, before a second compact brightening formed to the left of the slit. This feature moved into the axis of the slit (second panel; corresponding to the grouping of IRIS burst profiles beginning at $t\approx180$), then moved left out of the axis of the slit (third panel; leading to the detection of IRIS burst profiles stopping at $t\approx220$), then moved back into the axis of the slit (fourth panel; corresponding to the grouping of bursts starting at $t\approx250$), before finally moving right out of the axis of the slit (at $t\approx320$). During this time, this feature evolved dynamically, often rapidly increasing and decreasing in size. Finally, another compact brightening appeared along the slit (fifth panel; corresponding to the new grouping of IRIS burst profiles starting at $t\approx330$) before this moved right out of the line of the slit (sixth panel) where it remained until the end of the dataset. Despite the long duration of D1, the compact brightenings that occurred in this region had lifetimes of the order tens of minutes meaning each of these appeared to match the criteria for UV bursts (\citealt{Young18}).

\subsubsection{18 September 2014}

The final dataset studied here was sampled on $18$ September $2014$, over the course of slightly more than two hours. During this time, IRIS employed a sit-and-stare sequence with a large slit ($119$\arcsec\ slit length) centred on a plage region within AR $12166$ (roll angle of $20^\circ$). A total of $3190$ ($0.29$ \%\ of all pixels in this dataset) IRIS burst profiles were detected in this raster using our automated methods, with a temporal frequency of $0.42$ per second. In the top panel of Fig.~\ref{UVMax_fig6}, we once again plot a binary time-distance map displaying the locations of all IRIS burst profiles within this dataset. As with the datasets studied previously, the majority of the IRIS burst profiles identified here were isolated to one specific region (denoted by the box labelled `E1'). This region had a spatial length of $12$\arcsec\ along the slit and a temporal duration of $30$ minutes. Overall, $2909$ ($91.19$ \%) of the IRIS burst profiles identified in this dataset were contained within E1, with the other IRIS burst profiles predominantly being found at the same position along the slit, but at earlier times.

In the second row of Fig.~\ref{UVMax_fig6}, we plot this binary time-distance map purely for the region contained within E1. The IRIS burst profiles all occurred within an approximately $6$\arcsec\ region along the slit, with no velocities apparent along the plane-of-sky. The frequency with which these spectra were identified remained relatively consistently over a $25$ minute period, before reducing to zero. Once again, the vertical dashed red lines indicate the time-steps plotted for the SJI data (bottom row). Qualitatively comparing the locations of the IRIS burst profiles to the \ion{Si}{IV} $1394$ \AA\ line core (rest wavelength) intensity, we found similar results to the other previously discussed datasets, in that the IRIS burst profiles occurred close to, but not exclusively within, the brightest regions within the FOV during this time. 

\begin{figure*}
\includegraphics[width=0.99\textwidth]{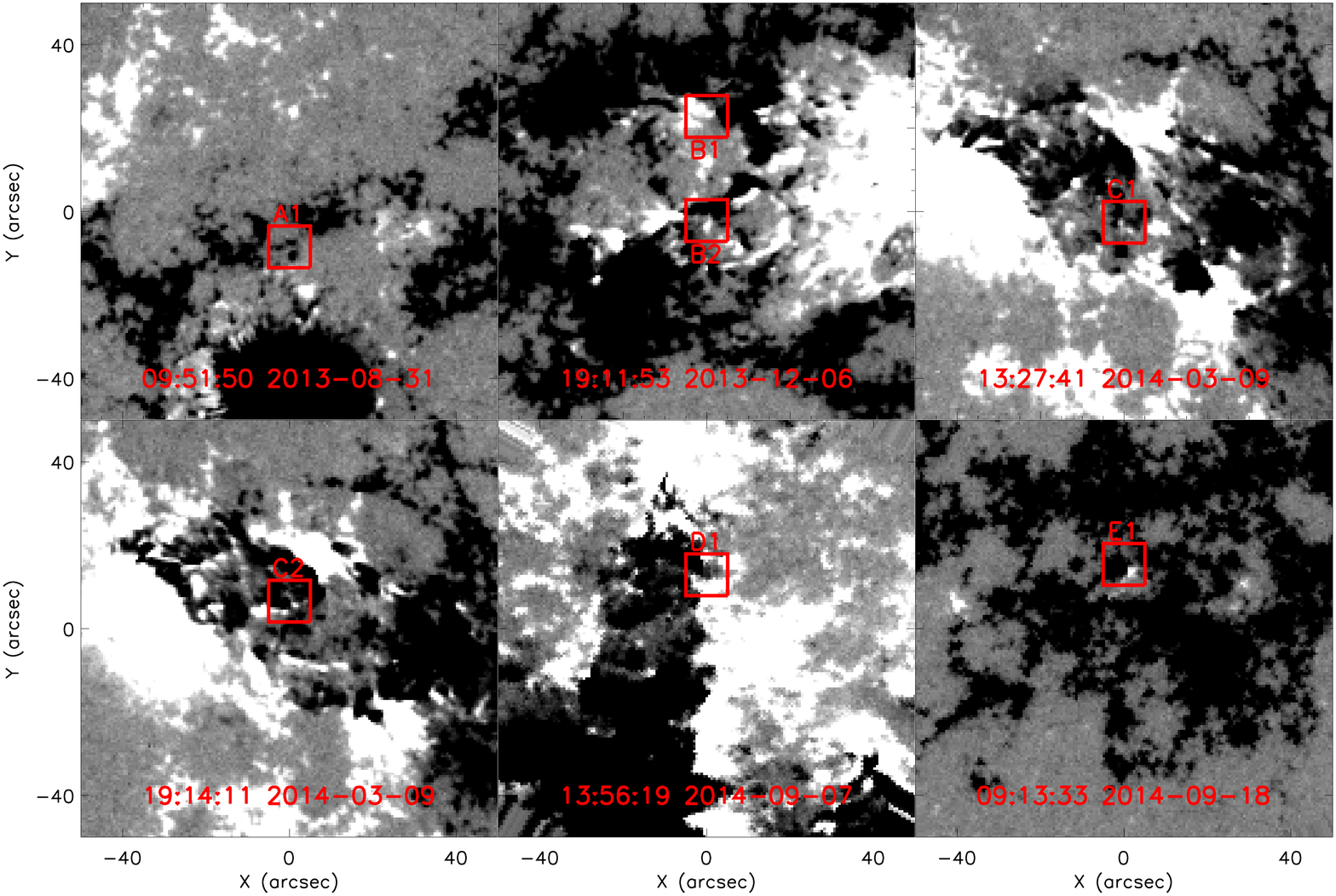}
\caption{SDO/HMI line of sight magnetic field maps saturated at $\pm100$ G for $100$\arcsec$\times100$\arcsec\ FOVs co-spatial to the IRIS SJI data. These maps were constructed using the SDO/HMI frame that was closest in time to the IRIS raster identified as the start-time of each of the seven regions discussed in detail here. The red labelled boxes indicate the approximate locations of each of these regions and outline the FOVs plotted at six time-steps in Fig.~\ref{UVMax_fig9}. Notably, B1 and B2 share a first time-step and, therefore, share a panel.}
\label{UVMax_fig8}
\end{figure*}

\begin{figure*}
\includegraphics[width=0.99\textwidth,trim={0 2cm 0 0}]{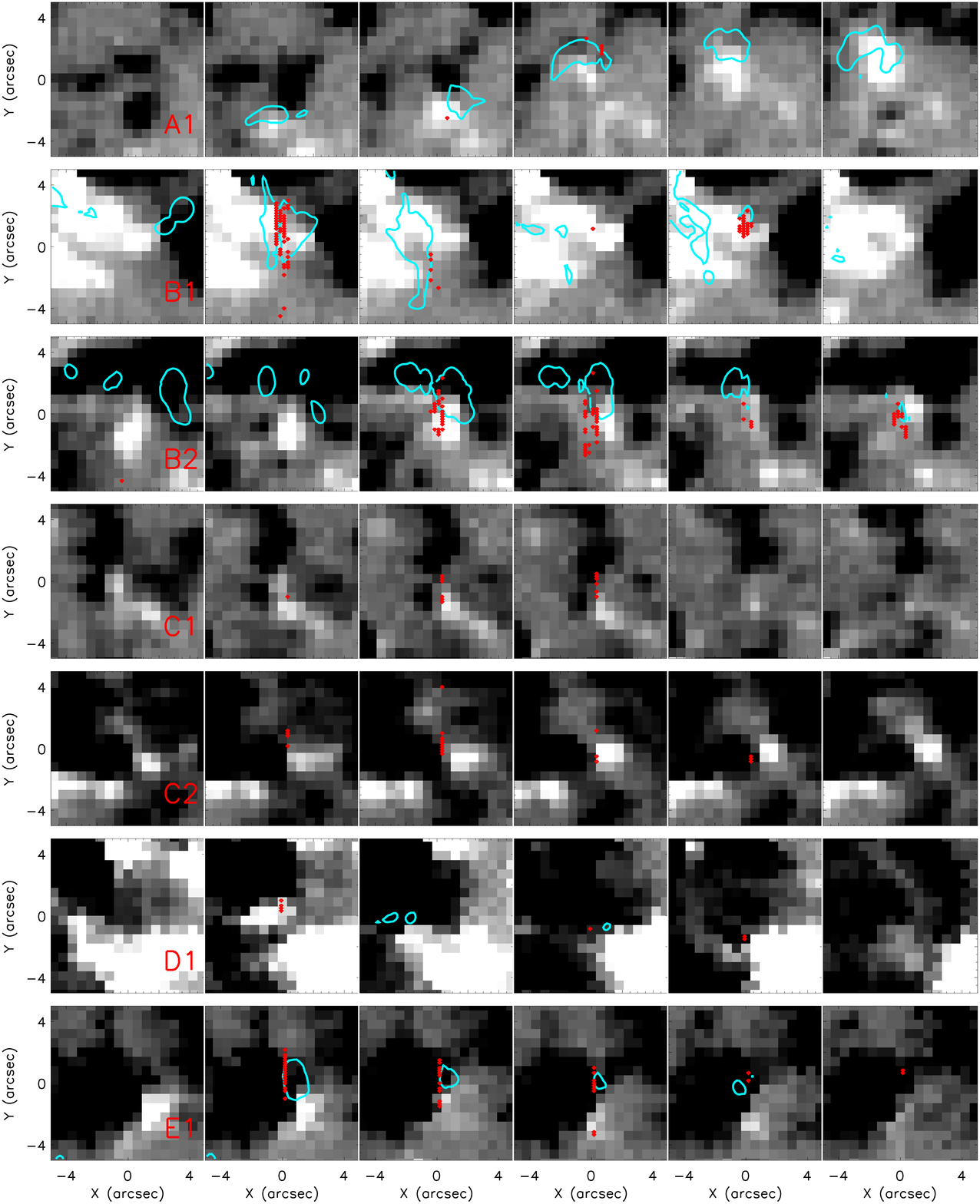}
\caption{Time series displaying the evolution of the line of sight magnetic field during the seven regions (with each row being labelled in the left-hand panel) that contained the majority of IRIS burst profiles identified here. The left-most panel corresponds to the first time-step of the region, while the right-most panel corresponds to the final time-step. The other panels were sampled at equal cadences between these two times. The red points indicate the approximate locations of the IRIS burst profiles identified by applying our automated methods to the temporally closest raster (A1 was observed by a two-step raster; B1 and B2 were observed by a four-step raster), whilst the aqua contours indicate the regions in the co-spatial and co-temporal IRIS $1400$ \AA\ SJI filter data that had intensities above $500$ DN/s. It is important to note that some panels contain no IRIS burst profiles. Opposite polarity magnetic fields were observed to interact throughout the duration of each of these regions.}
\label{UVMax_fig9}
\end{figure*}

Finally, we investigated the SJI data co-spatial to this region to better understand the dynamics occurring here (six time-steps are plotted in the bottom row of Fig.~\ref{UVMax_fig6}). Initially, a small compact brightening was evident to the right of the IRIS slit (first panel), before this moved into the axis of the slit and increased in size (second panel). This compact brightening then evolved dynamically over the next six minutes, decreasing in area before increasing in area again (third panel). Over the next $12$ minutes (fourth and fifth panels) the IRIS burst remained along the axis of the IRIS slit before it moved to the left (sixth panel) and faded from view. Notably, the two `islands' of IRIS burst profiles found at $Y\approx77$\arcsec\ were distinct from the main IRIS burst (as can be seen in the fourth panel), appearing as short jets with short (order of minutes) lifetimes. As with the previous features, this compact brightening in the SJI data satisfied the criteria that define UV bursts (\citealt{Young18}).

\subsection{\ion{Si}{IV} $1394$ \AA\ line core intensity of IRIS burst profiles}

The results of \citet{Kleint22} showed that IRIS burst profiles do not always display large increases in intensity at the \ion{Si}{IV} $1394$ \AA\ line core. On top of this, \citet{Nelson22} found that, on average, IRIS burst profiles have their peak intensity close to the rest wavelength of the line. However, our qualitative analysis of time-distance diagrams presented in the previous section indicated that many IRIS burst profiles did occur at least close to regions of increased intensity in the \ion{Si}{IV} $1394$ \AA\ rest wavelength (see Fig.~\ref{UVMax_fig5} for the clearest example). In order to investigate whether any quantitative relationship existed, we examined histograms of the \ion{Si}{IV} $1394$ \AA\ line core (rest wavelength) intensity for all pixels within any given raster as well as just those pixels where IRIS burst profiles were identified. In Fig.~\ref{UVMax_fig7}, we plot these histograms for each of the five datasets studied here. The larger top panels plot the histograms with the $y$-axis scaled appropriately for the entire raster (black bars), while the smaller bottom panels plot the histograms with the $y$-axis scaled appropriately for the IRIS burst profiles (red bars). Clearly, the IRIS burst profiles were not confined to the brightest pixels, with the peaks of the distributions for each of the five datasets being found close to $100$ DN/s.

Calculating the means of the distributions, we found that the average intensities of those pixels identified as IRIS burst profiles were higher (between $4$-$12$ times depending on the dataset) than the background; however, these were still below $300$ DN/s in each case. We note, though, that as we require enough intensity in the \ion{Si}{IV} $1394$ \AA\ line wings for absorption lines to be detected, it is possible that the differences in the means was purely due to our method, with low-intensity potential IRIS burst profiles being disregarded due to negligible line wing intensities. Finally, we also investigated the percentage of pixels with \ion{Si}{IV} $1394$ \AA\ line core (rest wavelength) intensities above $500$ DN/s (chosen as an arbitrary value for comparison) that displayed IRIS burst profiles for each dataset. We obtained highly variable results, with a value of $27.89$ \%\ found for the dataset sampled on $18$ September $2014$, compared to a value of $6.77$ \%\ for the dataset sampled on $9$ March $2014$. Overall, we found that the IRIS burst profiles studied here were not confined to only the brightest pixels, nor were they found in the majority of bright ($>500$ DN/s) pixels in any given dataset, despite often occurring close to regions of increased intensity (compared to the local background). These facts imply that our method of identifying IRIS burst profiles through the presence of super-imposed absorption lines on the \ion{Si}{IV} $1394$ \AA\ spectra alone is the most appropriate, rather than applying an arbitrary intensity threshold which would reject a high proportion of potential events. Essentially, we propose that although UV bursts are by definition brighter than the local background (\citealt{Young18}), IRIS burst profiles need not be.

\subsection{Properties of the co-spatial magnetic field}

In order to better contextualise our results, we also analysed the line of sight magnetic field co-spatial to the seven regions (in the time-distance domain, defined by the boxes over-laid on the upper maps plotted in Figs.~\ref{UVMax_fig2}-\ref{UVMax_fig6}) found to contain the majority of the IRIS burst profiles within these datasets. In Fig.~\ref{UVMax_fig8}, we plot $100$\arcsec$\times100$\arcsec\ FOVs sampled by the SDO/HMI instrument (saturated at $\pm100$ G) co-spatial to the IRIS SJI FOVs and co-temporal to the first time-step identified for each of the seven regions (B1 and B2 have the same first time-step and, hence, share a panel). The labelled $10$\arcsec$\times10$\arcsec\ red boxes indicate the approximate locations of the specific regions present within each panel. From these seven regions, five (B1, B2, C1, C2, and D1) occur in the centres of complex ARs containing highly structured opposite polarity magnetic fields, while two (A1 and E1) are found in seemingly less complex ARs that display only limited mixing of opposite polarities. The entire FOV surrounding E1 in particular, is almost uni-polar with only a small number of localised positive polarity (white pixels) islands amongst the negative polarity (black pixels) plage. As would be expected for events hypothesised to be driven by magnetic reconnection in the lower solar atmosphere, each of these seven regions were found to contain both positive and negative polarity magnetic fields at their initial time-step.

In Fig.~\ref{UVMax_fig9}, we plot the FOVs identified by the red boxes in Fig.~\ref{UVMax_fig8} at six different time-steps, with the first column corresponding to the start time of the respective region and the final column corresponding to the end time of the respective region. The red points indicate the positions of IRIS burst profiles identified in the temporally nearest raster while the aqua contours outline regions of increased intensity ($>500$ DN/s) within the co-spatial and co-temporal IRIS $1400$ \AA\ SJI data. Both the red points and the aqua contours are found close to inversion lines between the positive and negative polarity within the FOV. Additionally, it is immediately evident that a large number of IRIS burst profiles within these time-steps were detected close to but outside of the aqua contours, within the lower intensity background. Although the limited two-dimensional sampling of these rasters does not allow us to fully compare the relationship between IRIS burst profiles in the spectra and brightenings in the $1400$ \AA\ filter SJI data, the lack of one-to-one relationship between increased intensity and IRIS burst profiles does agree with the results presented in Fig.~\ref{UVMax_fig7}. 

Finally, investigating the evolution of the line of sight magnetic field throughout the evolution of each of these regions further confirmed that magnetic reconnection is a potential driver of the UV bursts in the SJI data and the IRIS burst profiles in the spectral data. For A1 (top row), the FOV was initially dominated by negative polarity magnetic field (first panel) before a small island of positive polarity moved into the FOV from the south (second panel), leading to cancellation of the negative polarity field (third, fourth, fifth, and sixth panels). Clear cancellation was also evident within these FOVs for B2 (third row), C1 (fourth row), and E1 (bottom row). For the remaining three FOVs (B1, C2, and D1), larger magnetic field elements dominated meaning obvious cancellation was not detected; however, the IRIS burst profiles were found close to the inversion lines between the opposite polarities. Notably, the plane-of-sky motions found in the locations of the IRIS burst profiles identified for A1 (second row of Fig.~\ref{UVMax_fig2}) and D1 (second row of Fig.~\ref{UVMax_fig5}) appear to correspond to the motions of the polarity inversion lines within these FOVs. The presence of such motions in data sampled by two different instruments further indicated that these motions were likely solar in nature (rather than caused by uncorrected wobbling from a single instrument).

\section{Conclusions}
\label{Conclusions}

In this article, we studied five datasets that were each found to contain more than $2000$ IRIS burst profiles by \citet{Kleint22}. These five datasets contributed a total of $16780$ (approximately $16.55$ \%) IRIS burst profiles to the total sample identified by those authors through analysis of the entire IRIS data catalogue sampled during $2013$ and $2014$ (see Fig.~\ref{UVMax_fig1}), meaning it is important to better understand the physical conditions at these locations. Using the algorithm as modified by \citet{Nelson22} lowered the number of returned IRIS burst profiles to $13904$. Our research into the IRIS burst profiles returned by the new algorithm in these five datasets found that:
\begin{itemize}
\item{The majority ($89.19$ \%) of all IRIS burst profiles found within these datasets were confined within seven small regions in the time-distance domain (temporal durations of $<4$ hours and distances of $<12$\arcsec\ along the IRIS slit; see Figs.~\ref{UVMax_fig2}-\ref{UVMax_fig6}). Analysis of IRIS SJI data co-spatial and co-temporal to these regions indicated that features matching the defining characteristics of UV bursts (as listed by \citealt{Young18}) formed and evolved along the axis of the IRIS slit during these times. The UV bursts were relatively long-lived (lifetimes of $>10$ minutes) and sometimes repetitive over the course of several hours (see, for example, Fig.~\ref{UVMax_fig5}).}
\item{IRIS burst profiles were qualitatively found close to both regions of increased intensity in \ion{Si}{IV} $1394$ \AA\ intensity maps and the UV bursts found in the IRIS $1400$ \AA\ SJI data. When this potential relationship was examined in more detail using quantitative methods (see Fig.~\ref{UVMax_fig7}), however, it was found that the majority of IRIS burst profiles occurred in pixels with relatively low \ion{Si}{IV} $1394$ \AA\ line core (rest wavelength) intensities ($<500$ DN/s), implying the presence of a large number of spectra with super-imposed absorption lines outside the cores of the UV bursts apparent in the SJI data. Additionally, IRIS burst profiles did not account for the majority of bright pixels ($>500$ DN/s) in any dataset, with values ranging from $6.77$ \%\ to $27.89$ \%. Essentially, this implies that the use of an arbitrary intensity threshold will cause any IRIS burst profile detection algorithm to miss a large proportion of these events. In addition to this, we propose that the signatures of magnetic reconnection in spectral data occur over a larger area than the signatures apparent in imaging data.}
\item{Finally, we examined the structuring and evolution of the line of sight magnetic field co-spatial to the seven regions containing the majority of IRIS burst profiles. Five of these regions (B1, B2, C1, C2, and D1) occurred within three highly structured ARs, while the other two (A1 and E1) were found in two less complex ARs (see Fig.~\ref{UVMax_fig8}). This implies that the presence of large numbers of IRIS burst profiles in a single dataset does not indicate any specific level of complexity for the observed AR. Opposite polarity magnetic fields were present within $10$\arcsec$\times10$\arcsec\ boxes surrounding these IRIS burst profiles during the times when they were observed, with four regions (A1, B2, C1, and E1) displaying clear evidence of cancellation through time and the other three (B1, C2, and D1) containing large, opposite polarity magnetic elements. This indicates that magnetic reconnection could be the predominant driver of these IRIS burst profiles (see Fig.~\ref{UVMax_fig9}). We do note that, of course, cancellation in the solar photosphere can occur without co-spatial and co-temporal IRIS burst profiles being present.}
\end{itemize}
Overall, the majority of IRIS burst profiles are linked to a small number of UV bursts, which by chance were sampled by the IRIS slit for their entire lifetimes, within otherwise unexceptional host ARs. We are, therefore, currently unable to predict whether any given dataset will contain a large number of IRIS burst profiles.

\begin{acknowledgements}
CJN is thankful to ESA for support as an ESA Research Fellow. LK gratefully acknowledges funding via a SNSF PRIMA grant. IRIS is a NASA small explorer mission developed and operated by LMSAL with mission operations executed at NASA Ames Research Center and major contributions to downlink communications funded by ESA and the Norwegian Space Centre. SDO/HMI data provided courtesy of NASA/SDO and the HMI science team. This research has made use of NASA’s Astrophysics Data System Bibliographic Services.
\end{acknowledgements}

\bibliographystyle{aa}
\bibliography{UV_Max}

\end{document}